\documentclass[ reprint, aps, prl, two column, showkeys]{revtex4-1}
\usepackage{graphicx,epsfig}
\usepackage{amssymb}
\usepackage{amsmath}
\usepackage{bm}
\usepackage{textcomp}
\usepackage{soul,xcolor}
\usepackage{float}

\usepackage{color}

\begin{document}
\setcounter{page}{1}
\setstcolor{red}

\title[]{Understanding limits to mobility in ultra-high-mobility GaAs two-dimensional electron systems: The quest for 100 million cm$^2$/Vs and beyond}
\author{Yoon Jang \surname{Chung}}
\affiliation{Department of Electrical Engineering, Princeton University, Princeton, NJ 08544, USA  }
\author{A. Gupta}
\affiliation{Department of Electrical Engineering, Princeton University, Princeton, NJ 08544, USA  }
\author{K. W. \surname{Baldwin}}
\affiliation{Department of Electrical Engineering, Princeton University, Princeton, NJ 08544, USA  }
\author{K. W. \surname{West}}
\affiliation{Department of Electrical Engineering, Princeton University, Princeton, NJ 08544, USA  }
\author{M. \surname{Shayegan}}
\affiliation{Department of Electrical Engineering, Princeton University, Princeton, NJ 08544, USA  }
\author{L. N. \surname{Pfeiffer}}
\affiliation{Department of Electrical Engineering, Princeton University, Princeton, NJ 08544, USA  }

\date{\today}

\begin{abstract}
For several decades now, ultra-high-mobility GaAs two-dimensional electron systems (2DESs) have served as the hallmark platform for various branches of research in condensed matter physics. Fundamental to this long-standing history of success for GaAs 2DESs was continuous sample quality improvement, which enabled scattering-free transport over macroscopic length scales as well as the emergence of a diverse range of exotic many-body phenomena. While the recent breakthrough in the quality of GaAs 2DESs grown by molecular beam epitaxy is highly commendable in this context, it is also important and timely to establish an up-to-date understanding of what obstructs us from pushing the mobility limit even further. Here, we present mobility data taken at a temperature of 0.3 K for a wide variety of state-of-the-art GaAs 2DESs, exhibiting a maximum, world-record mobility of $\mu\simeq57\times10^6$ cm$^2$/Vs at a 2DES density of $n=1.55\times10^{11}$ /cm$^2$. We also provide comprehensive analyses of the collective scattering mechanisms that can explain the results. Furthermore, based on our study, we discuss potential scenarios where GaAs 2DES mobility values exceeding $100\times10^6$ cm$^2$/Vs could be achieved.


\end{abstract}
\maketitle

\section{I. Introduction}

Few experimental platforms in condensed matter physics can match the breadth and depth offered by low-disorder two-dimensional electron systems (2DESs). From everyday field-effect transistors to the more intricate electronic devices that require ballistic or quantum coherent transport, applications that utilize 2DESs are widespread in the scientific community. While essential to each of these cases, there is one particular circumstance in which 2DESs pose a unique opportunity: the study of electron-electron interaction. At cryogenic temperatures where the kinetic energy of a 2DES is determined by the Fermi energy, the carrier density of a 2DES can be controlled so that the Coulomb energy becomes comparatively dominant. A magnetic field perpendicular to the 2DES can further enhance this inclination via Landau quantization of the density of states.

Several many-body phases have emerged in 2DESs over the past few decades. Notable examples include the odd- and even-denominator fractional quantum Hall states (FQHSs) \cite{Tsui,Willett}, stripe/nematic phases \cite{stripe1,stripe2}, Wigner solids \cite{HeWigner,Wigner1,Wigner2,Wigner3}, and Bose-Einstein exciton condensates \cite{B1,B2,B3}. In most cases, the earliest experimental observations of such exotic states were made in 2DESs hosted in GaAs quantum wells (QWs) grown by molecular beam epitaxy (MBE) (for reviews, see \cite{ShayeganReview,JainBook,HalperinBook}. It is crucial that disorder is minimized for delicate electron-interaction phenomena to develop without hindrance, and no other system has progressively improved in terms of sample quality as much as the GaAs/AlGaAs materials group \cite{EnglishM,ParisM,ShayeganM,ShayeganM2,modulation,Stormer,PfeifferM,LorenPhysica,Umansky35,Schlom,Manfra35,HighMobility,HighHole}. History demonstrates that numerous unexpected interaction-driven phases materialize with better GaAs 2DES quality, which provides a strong incentive for continuous advancement on this front.

Oftentimes, the quality of a 2DES is evaluated by measuring the transport mobility ($\mu$) of the sample. In a simple Drude model, $\mu$ is directly proportional to the scattering lifetime ($\tau$) via the relation $\mu=e\tau/m^*$, where $e$ is the fundamental charge and $m^*$ is the effective mass of the electrons in the 2DES. Within the same material system, a higher $\mu$ hence implies a larger $\tau$, meaning that an electron can travel in the 2DES without experiencing a scattering event for a longer time. Multiple studies have reported in-depth analysis on what limits $\mu$ in ultra-high-quality GaAs 2DESs, issuing directions for further sample quality improvement \cite{EnglishM,ParisM,ShayeganM,ShayeganM2,modulation,Stormer,PfeifferM,LorenPhysica,Umansky35,Schlom,Manfra35,HighMobility,HighHole,Jiang,Gold,DS1,Manfra35,DS2,Boris1,Boris2,DS3,Alproblem,Dopingwell,Gold2,Boris3,WegS}. Following the recent breakthrough in GaAs 2DES mobility up to $\mu\simeq44\times10^6$ cm$^2$/Vs \cite{HighMobility}, we provide a timely update to these works and report a world-record mobility of $\mu\simeq57\times10^6$ cm$^2$/Vs at $T=0.3$ K in a 38.5-nm-wide GaAs QW hosting a 2DES with a density of $n=1.55\times10^{11}$/cm$^2$. At higher densities, however, the mobility begins to drop. We review the various scattering mechanisms that contribute to determining $\mu$ in state-of-the-art samples. A careful comparison is made between our models and experimental data, based on which we suggest guidelines for achieving $100\times10^6$ cm$^2$/Vs mobility and beyond in the future.

 \begin{figure*}[t]
\centering
    \includegraphics[width=.98\textwidth]{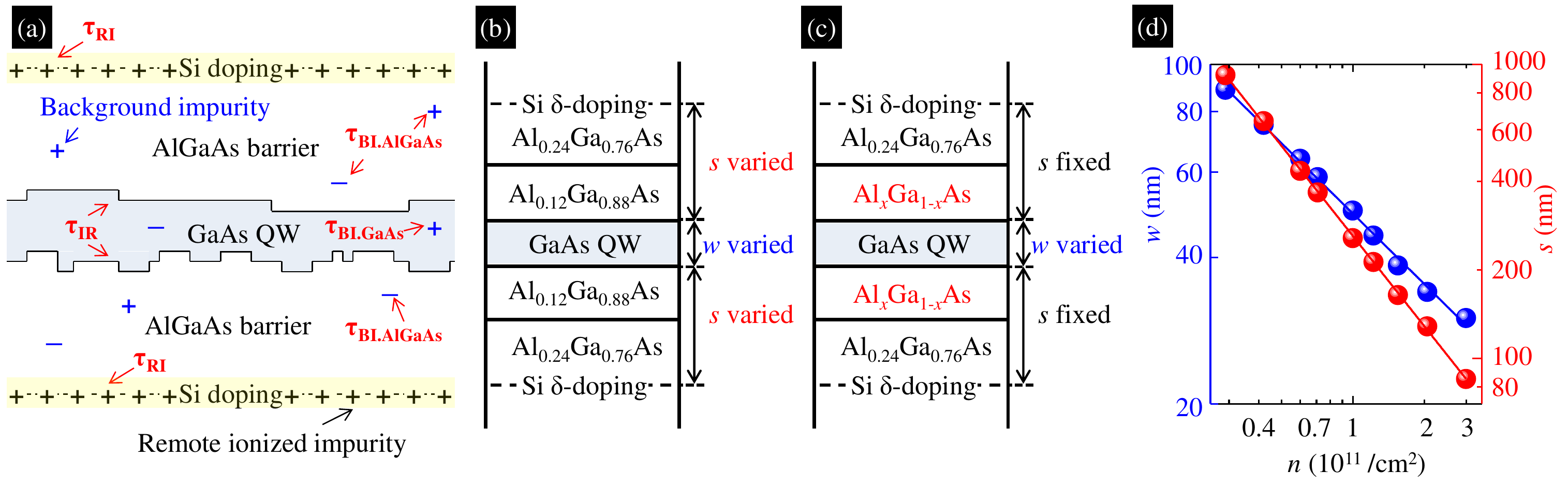}

 \caption{\label{fig1} (a) A schematic diagram depicting the various scattering mechanisms for 2DESs hosted in modulation-doped GaAs QWs. The scattering times from background impurities in the GaAs channel ($\tau_{BI.GaAs}$) as well as the AlGaAs barrier ($\tau_{BI.AlGaAs}$), remote ionized impurities that are generated by doping ($\tau_{RI}$), and layer fluctuation driven interface roughness ($\tau_{IR}$) all contribute to the total amount of scattering and are each shown in red. The charged species from residual background impurities are marked blue while those from intentional ionized dopants are marked black. (b) The layer stack structure for the series of samples that we use to evaluate the $\mu$ vs. $n$ behavior of ultra-high-quality GaAs 2DESs. The Al alloy fractions ($x$) of the Al$_x$Ga$_{1-x}$As stepped-barrier stack near the QW are fixed, and the spacer thickness ($s$) is varied to control $n$ while the QW width ($w$) is also varied accordingly so that the second electric subband is not occupied in higher density samples. (c) The layer stack structure that we use to evaluate the $\mu$ vs. $w$ behavior of ultra-high-quality GaAs 2DESs. In this case, $s$ is fixed for a given $n$, but $w$ is varied for a series of samples with different barrier alloy fractions near the QW. Note that the Si doping shown in (a)-(c) is only schematic; in all of our samples we used the doping-well scheme (see text). (d) Shows the $s$ and $w$ values as a function of $n$ for the samples with the structure in (b).}
\end{figure*}	

\section{II. Scattering mechanisms in ultra-high-quality GaAs 2DES\MakeLowercase{s}}

Any circumstance that causes a traversing electron in the 2DES to partially or fully lose its externally set initial information can be classified as a scattering event. There are several mechanisms that can give rise to such an occurrence. For instance, charged impurities near or in the GaAs QW where the 2DES resides would locally alter the potential landscape for electrons and act as scattering centers. Figure 1 (a) schematically summarizes the typical scattering mechanisms considered when analyzing ultra-high-quality GaAs 2DESs. It is useful to associate a characteristic $\tau_i$ with each scattering mechanism for discussion throughout the rest of the paper. The holistic scattering time $\tau$ can then be determined from Mathiessen's rule $1/\tau=\Sigma \frac{1}{\tau_i}$. 

We start by defining $\tau_i$ terms that are linked with scattering from charged impurities. Unless grown from 100\% pure material and in absolute vacuum with perfectly inert heating sources and chamber components, a finite amount of residual background impurities are unavoidable in samples. When charged, these impurities act as scattering centers, as alluded to in the previous paragraph. The impact of such scattering centers on electronic transport in the sample depends on their proximity to the 2DES. Impurities in the GaAs QW are the strongest scatterers, and the influence gradually weakens as the charged species are moved further away from the GaAs channel. In this context, for GaAs 2DESs hosted in MBE grown QWs, as shown in Fig. 1(a) we group scattering events and hence $\tau$ from background impurities into two categories \cite{Jiang,Gold,Gold2}: scattering from background impurities in the GaAs channel ($\tau_{BI.GaAs}$) and in the AlGaAs barrier ($\tau_{BI.AlGaAs}$).     

Another part of ultra-high-quality GaAs samples where charged impurities are inevitable is the dopant layer. Modulation doping is standard for state-of-the-art structures, and the presence of remote ionized Si impurities is necessary to host a 2DES in the GaAs QW in this scheme \cite{modulation,Stormer}. In contrast to the case of residual background impurities where a scarce amount is spread out ubiquitously, remote ionized impurities are typically highly concentrated in a plane positioned a certain distance away from the GaAs channel (see Fig. 1(a)). This is because in ultra-high-quality GaAs 2DES samples, the Si dopants are introduced to the sample in a $\delta$-function-like manner per design \cite{Delta,Ploog,Si1,Si2,Si3,Schubert}, where the spacer thickness ($s$) between the dopant layer and the GaAs QW determines the 2DES density \cite{Davies,Designrules}. The densely-packed nature of remote ionized impurities suggests that the scattering time deriving from them ($\tau_{RI}$) could contribute significantly to the holistic $\tau$, especially when $s$ is small or when other $\tau_i$ terms become relatively large.

Finally, we also take into account the scattering time associated with interface roughness ($\tau_{IR}$). It is well established that layer fluctuations occur during the MBE growth of GaAs/AlGaAs heterostructures \cite{Ourmazd,Salemink,Zrenner}. While thickness variation on the order of a few monolayers seems inconsequential in most cases, it could have serious repercussions when it occurs at the GaAs QW/barrier interface, as shown in Fig. 1(a). Layer fluctuations in this region of the structure, or `interface roughness', causes the electron energy and charge distribution to vary in the 2DES channel. This abrupt change in the local 2DES density acts as a scattering potential for electrons, and its effect has been studied in detail both theoretically \cite{Ando,GoldIR1,GoldIR2,Li} and experimentally \cite{Sakaki,Doby}. 

The goal of this paper is to investigate the influence of each of the scattering terms outlined above on the mobility of state-of-the-art, ultra-high-quality GaAs 2DESs. By establishing a thorough understanding of what limits quality in the best available samples, we hope to provide a basis to strategize the next steps forward.   


\section{III. Details of sample structures}

Two different series of samples were grown for our study, as schematically shown in Figs. 1(b) and (c). The first set of samples is designed to explore the change in mobility of ultra-high-quality GaAs 2DESs as a function of electron density, $n$. This series of samples aims to provide a broad sketch of the impact of different scattering terms on the mobility. Figure 1(b) depicts the layer stack structure of such samples, where the Al$_x$Ga$_{1-x}$As spacer layer thickness ($s$) is varied to control $n$. For each sample, the QW width ($w$) is also varied so that it is sufficiently wide to minimize interface roughness scattering, but not so wide that electrons start populating the second electric subband. Figure 1(d) summarizes the experimental parameters of these samples. To ensure ultra-high-quality, all samples here implement a stepped-barrier structure, where the barrier alloy fraction $x$ is graded down from $x=0.24$ near the $\delta$-layer Si dopants to $x=0.12$ near the GaAs/Al$_x$Ga$_{1-x}$As interface. 


The second series of samples, shown in Fig. 1(c), focuses on examining the effect of interface roughness scattering in ultra-high-quality GaAs 2DESs. We achieve this by varying $x$ at the GaAs/Al$_x$Ga$_{1-x}$As interface and $w$ at a fixed $n$, and measuring the mobility of each sample. At a given $x$, several samples with a range of $w$ were grown to probe the evolution of mobility with $w$ for two different electron densities. It is worth emphasizing that the Si doping we show in Figs. 1(a)-(c) is only schematic. For all the samples reported here, we used a doping-well scheme where dopants are introduced into a narrow GaAs QW flanked by AlAs QWs on both sides; for details, see Refs. \cite{Boris1,Boris2,Dopingwell,HighMobility,Manfra35}. 

In the following sections of the paper, we will discuss and analyze the data from all of our samples in detail and attempt to explain them in accordance with the scattering mechanisms described in Section II.


\begin{figure*}[t]
 
 \centering
    \includegraphics[width=.98\textwidth]{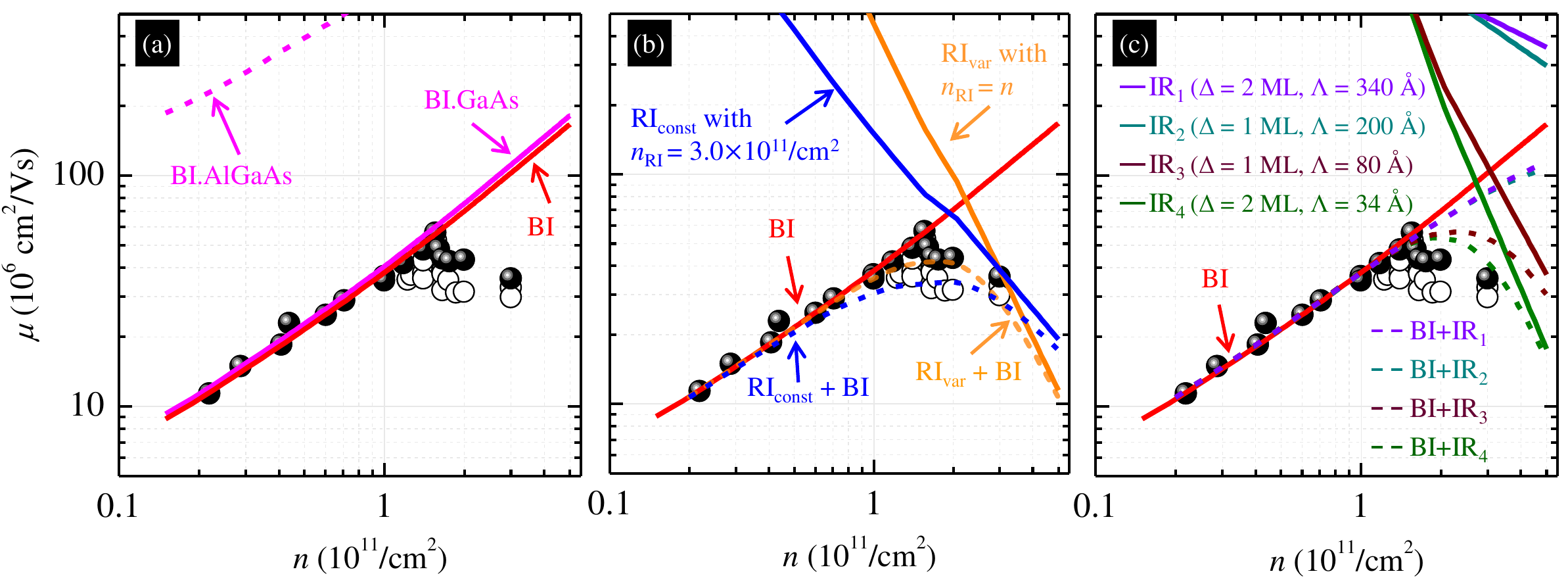} 
  \caption{\label{fig2} Experimental $\mu$ vs. $n$ data ($T=0.3$ K) for our ultra-high-quality GaAs 2DESs and their comparison with simple models. Multiple samples were grown and measured for the same density, and the closed black circles denote the highest mobility observed at each density. All lower mobility values are shown with open black symbols. The curves represent the $\mu$ vs. $n$ trends expected from various scattering mechanisms. The solid and dashed magenta curves in (a) show the expected evolution of $\mu$ vs. $n$ solely based on $\tau_{BI}$, assuming background impurity concentrations of $1.25\times10^{13}$ and $2.5\times10^{13}$ /cm$^3$ for the GaAs QW and AlGaAs barriers, respectively. The solid red curve marked BI shows $\mu$ deduced from the sum of these two $\tau_{BI}$ contributions based on Matthiessen's rule. In (b) we display the $\mu$ vs. $n$ trends based on $\tau_{RI}$, with the blue solid curve showing the results for a fixed total remote impurity density of $n_{RI}=3.0\times10^{11}$ /cm$^2$ and the orange solid curve for when $n_{RI}$ is equal to the 2DES density $n$; the red curve is the same as in (a), representing the mobility limited by background impurities only. The kink observed in the blue and orange solid curves for remote ionized impurity scattering derives from the fact that, in our calculations, we use the actual $w$ values of our samples, which do not fall exactly on a power law curve (see Fig. 1(d)). The dashed blue and orange curves correspond to $\mu$ values deduced from Matthiessen's rule considering both $\tau_{BI}$ and $\tau_{RI}$ for each case. Similarly, the solid curves in (c) show calculated $\mu$ vs. $n$ trends for various $\tau_{IR}$ conditions. The $\Delta$ and $\Lambda$ values used for the different colored curves are given in the legend of the figure. }
\end{figure*} 

\section{IV. Mobility vs. 2DES density: Experimental data}

	Figure 2 shows the measured 2DES density vs. mobility data of our samples. Every data point comes from a separate wafer piece evaluated in the van der Pauw geometry using low-frequency lock-in amplifiers at $T=0.3$ K. Eutectic In/Sn is placed on each of the four corners and flats of 4 mm$\times$4 mm samples and annealed at $T=425$ \textdegree C in an Ar:H$_2$ (95\%:5\%) environment to establish 8 electronic contacts to the 2DES. For measurement, our samples are loaded into a $^3$He cryostat, where a red light-emitting diode (LED) is used to illuminate the samples for 5 minutes at $T\sim10$ K. Following illumination, we wait for 30 minutes at $T\sim10$ K after the LED has been turned off before resuming the cool down to base temperature. The 2DES density $n$ is then determined by assigning a value that is concomitant with the quantum Hall features observed in the magnetoresistance trace. The mobility is deduced from the Drude model, $\mu=1/\rho ne$, where $\rho=\pi R_{ave}/ln(2)$ with $R_{ave}$ being the average value of resistance measured from all possible four-probe contact configurations in the sample. We find that for a given sample the 2DES density and mobility typically vary by less than 5\% when repeating the procedure outlined above, even if it has experienced a full thermal cycle to room temperature. For some of the cases in Fig. 2, multiple samples with an identical layer structure but from different growths are measured and plotted for the same $n$. The highest mobility measured from such campaigns are highlighted with solid black circles, while other attempts are represented by the open black circles. Throughout this work, we focus on the highest mobility results shown as the solid black circles when comparing the experimental data with calculations from our models.  	
	
	The peak mobility in Fig. 2 generally increases with $n$ and reaches a maximum value of $\mu\simeq57\times10^6$ cm$^2$/Vs at $n=1.55\times10^{11}$/cm$^2$, but then drops at higher $n$. Before we set up the models to understand this behavior and examine the influence of various scattering mechanisms on the mobility of our GaAs 2DESs, we reiterate that each of the data points in Fig. 2 come from different samples with distinct structural parameters such as QW width $w$ and spacer thickness $s$ as mentioned in Section III. For example, the samples with $n\simeq2.2\times10^{10}$, $6\times10^{10}$, $1.2\times10^{11}$, and $3\times10^{11}$ /cm$^2$ have $w$ values of 95, 64, 44, 30 nm and $s$ values of 1282, 434, 213, and 80 nm, respectively (see Fig. 1(d)). The variation in these parameters can have a significant impact on scattering, and we incorporate 2DES density specific $w$ and $s$ values commensurate with our samples when calculating the mobility values discussed in Sections V and VI.

\section{V. Mobility vs. 2DES density: Calculations}

	The electron density plays an important role in determining scattering times in 2DESs. This becomes apparent when considering that the scattering rates deduced from Fermi's golden rule are functions of the Fermi wavevector $k_F$, which is inherently dependent on $n$. When $k_F$ changes, not only does the phase space available for scattering change, but so does the screening behavior of the 2DES. Even in the simple, single-particle-based, zero-temperature model, the effect of varying $k_F$ could be profound, which warrants thorough analyses. Using this framework, we carefully examine how each of the scattering terms are expected to behave as a function of $n$ and compare our findings with the experimental data.


\subsection{1. Residual background impurity scattering}	

	It is unrealistic to assume that there are absolutely no residual impurities in a GaAs 2DES sample. If any of the impurities are charged, they will displace the trajectory of electrons traversing in the vicinity. While all scattering events `erase' the memory of an incoming electron, some are less detrimental to the scattering times that are relevant to the transport mobility. For example, the complete backscattering of an electron should impede current flow substantially more compared to a small-angle scattering event. Indeed, formulations that are based on this assumption have been widely implemented to interpret transport scattering lifetimes and mobility over a wide range of situations \cite{Davies,DS1,DS2,DS3,Boris1,Boris2,Boris3,Gold,Gold2,GoldIR1,GoldIR2,Jiang}. Supposing a 2D sheet of impurities with a concentration of $n_{imp}$ placed a distance $z$ from the 2DES, typically such expressions have the form \cite{Davies}:
	
\begin{equation}
   \frac{1}{\tau}=n_{imp}F(k_F)\int_{0}^{2k_F} \frac{e^{-2q|z|}}{(q+q_{TF})^2}\frac{q^2\,dq}{\sqrt{1-(q/2k_F)^2}}.
\label{eqn:1}
 \end{equation}
Here, $q_{TF}$ is the Thomas-Fermi wavevector and $F(k_F)=\frac{m^*}{2\pi\hbar^3 k_F^3}(\frac{e^2}{2\epsilon_0\epsilon_b})^2$, where $\hbar$ is the reduced Planck's constant, $\epsilon_0$ the vacuum permittivity, and $\epsilon_b$ the dielectric constant of the background material ($\epsilon_b\simeq13$ for GaAs and Al$_x$Ga$_{1-x}$As with small $x$). 

	Considering the sample structure of the quintessential ultra-high-quality GaAs 2DESs, when using the model described by Eq. (1) it is useful to differentiate between two regions of the sample where residual background impurities exist; the GaAs QW that hosts the 2DES, and the AlGaAs barrier that flanks it. This is because usually Al is regarded as a getter metal while Ga is not \cite{Alproblem}, implying that under similar purification and vacuum conditions the AlGaAs barrier could have a higher impurity concentration compared to the GaAs QW. Another beneficial aspect of separately analyzing the barrier and the QW emerges when taking the finite (non-zero) thickness of the electron wavefunction into account, which adds significant complexity to the problem for the QW region \cite{Ando,Davies,Gold}. For the sake of simplicity, in this paper we neglect such corrections.
	
	
	Under this scheme, it is relatively straightforward to modify Eq. (1) to find an expression for the scattering rate in the GaAs QW. Presuming that the background impurity concentration in the GaAs QW ($n_{BI.GaAs}$) is constant, we can relate it to $\tau_{BI.GaAs}$ as:
	
\begin{equation}
\begin{split}
   \frac{1}{\tau_{BI.GaAs}}&=n_{BI.GaAs}F(k_F)
\\   
   &\times\int_{0}^{2k_F}\!\!\!\! \int_{-w/2}^{w/2}\frac{e^{-2q|z|}\,dz}{(q+q_{TF})^2}\frac{q^2\,dq}{\sqrt{1-(q/2k_F)^2}}.
\label{eqn:2}
\end{split}
 \end{equation}
 Here we are merely integrating scattering contributions from infinitesimal 2D sheets of background impurities over the width of the GaAs QW, based on Eq. (1).

	Similarly, we can estimate $\tau_{BI.AlGaAs}$ assuming a constant background impurity concentration in the AlGaAs barrier ($n_{BI.AlGaAs}$). We find:
\begin{equation}
\begin{split}
   &\frac{1}{\tau_{BI.AlGaAs}}=n_{BI.AlGaAs}F(k_F)
 \\
   &\times2\int_{0}^{2k_F}\!\!\!\! \int_{w/2}^{\infty}\frac{e^{-2qz}\,dz}{(q+q_{TF})^2}\frac{q^2\,dq}{\sqrt{1-(q/2k_F)^2}},
\label{eqn:3}
\end{split}
 \end{equation}
The factor of 2 in front of the integral comes from the fact that the GaAs 2DES is flanked by AlGaAs barriers on both sides, and we assume that the thickness of the barrier is infinite. This is a reasonably fair approach since the exponential decay with $z$ truncates the influence from barrier regions further away from the 2DES. Indeed, we compute that contributions from layers with $z\gtrsim200$ nm have minimal impact on the scattering rate.  
 
	By inputting sample parameters to Eqs. (2) and (3), we can compare the results to experimental data using the relation $\mu=e\tau/m^*$ based on the Drude model to estimate $n_{BI.GaAs}$ and $n_{BI.AlGaAs}$. Figure 2(a) shows the $\mu$ vs. $n$ data of our ultra-high-quality GaAs 2DES, juxtaposed with profiles generated for scattering in the GaAs QW (solid magenta curve) and the AlGaAs barrier (dashed magenta curve) assuming impurity concentrations of $1.25\times10^{13}$ /cm$^3$ and $2.5\times10^{13}$ /cm$^3$, respectively. The $\mu$ vs. $n$ trend that is obtained from the collective scattering of these two terms via Mathiessen's rule ($1/\tau_{BI}=1/\tau_{BI.GaAs}+1/\tau_{BI.AlGaAs}$) is shown as the solid red curve marked $\tau_{BI}$. 
	
	



	It is clear from Fig. 2(a) that there is good agreement between our data and what is expected from Eqs. (2) and (3) when $n\lesssim1.6\times10^{11}$/cm$^2$. For the sample structure we use in this work, in this low density regime, the spacer is moderately thick ($s\gtrsim180$ nm) meaning that the remote ionized impurities are a substantial distance away from the 2DES, so that $\tau_{RI}$ should not have a large effect on the mobility. Furthermore, the low density allows proportionately wide QWs to host the 2DES while still avoiding second subband occupation, which implies that interface roughness scattering is also minimized. From these perspectives, it then seems reasonable that the results deduced from background impurity scattering models coincide well with the experimental data at small $n$. Similar arguments have been made for low-density, modulation-doped GaAs 2DES where background impurity scattering is expected to be the primary factor in determining the mobility \cite{Gold,Jiang,ShayeganM,PfeifferM,Gold2,DS1,DS2,DS3,Boris3,HighMobility}.  
	
\subsection{2. Remote ionized impurity scattering}	

	As mentioned briefly earlier, double-sided, modulation-doped structures are typical for ultra-high-quality GaAs 2DESs. Such sample design requires a sheet of ionized impurities at a distance $s$ away from the 2DES on both sides of the QW \cite{modulation,Stormer,Delta,Ploog,Schubert,Davies,Designrules,Si1,Si2,Si3,HighMobility,EnglishM,ShayeganM,ShayeganM2,PfeifferM,Umansky35,Manfra35,ParisM,LorenPhysica,Schlom}. Scattering from these inevitable charges can be estimated by evaluating a simpler version of Eq. (3), where we only consider an isolated layer of impurities rather than a uniform bulk distribution:
\begin{equation}
\begin{split}
   \frac{1}{\tau_{RI}}=n_{RI}F(k_F)\int_{0}^{2k_F}\frac{e^{-2q(s+w/2)}}{(q+q_{TF})^2}\frac{q^2\,dq}{\sqrt{1-(q/2k_F)^2}}.
\label{eqn:4}
\end{split}
 \end{equation}
Here, $n_{RI}$ is the total sheet density of the remote ionized impurities from both sides of the QW. Unlike residual background impurities which are inadvertently accumulated in the sample during the growth process, remote ionized impurities are intentionally introduced to the structure in a controlled fashion to act as dopants that generate the 2DES. Assuming that all the 2D electrons come from modulation doping, charge neutrality implies that at least one ionized dopant is needed for every carrier in the channel. It is then appropriate to assume that $n_{RI}\geq n$, since it may be necessary for an additional number of dopants to be ionized to compensate for the charged defects throughout the structure. 
	As shown in Fig. 2, the 2DES density of the samples studied in this section range from $2.2\times10^{10}$ /cm$^2$ to $3.0\times10^{11}$ /cm$^2$. Based on the discussion from the previous paragraph, we consider two different $n_{RI}$ values for the analysis of $\tau_{RI}$; in one case we take $n_{RI}=3.0\times10^{11}$ /cm$^2$ so that it is constant and corresponds to the highest 2DES density of our samples (solid blue curve in Fig. 2(b)), and in another we take it to be variable and equal to the 2DES density ($n_{RI}=n$, solid orange curve in Fig. 2(b)). Using Matthiessen's rule to also include all scattering contributions from the residual background impurities as deduced earlier, it seems that both assumptions agree reasonably well with our data (see the dashed blue and orange curves in Fig. 2(b)).
	
	It is important to note here that, unlike many previous reports where the GaAs 2DES density is varied in a structure with a fixed spacer thickness using illumination or a gate voltage \cite{Jiang,ShayeganM,PfeifferM}, here we are evaluating samples with different spacer thicknesses for each 2DES density (see Fig.1 (d)). The spacer thickness must be decreased to obtain a higher electron density. The decreasing trend of $\mu$ vs. $n$ for remote ionized impurity scattering that we see in Fig. 2(b) reflects the fact that bringing a sheet of impurities closer to the 2DES overcomes the benefits of the better screening provided by the increased $n$ in our sample structures. This tendency is stronger when $n_{RI}=n$ compared to when $n_{RI}$ is constant since higher density samples then entail an increasing density of sheet impurities as they are moved closer and closer toward the QW. Such differentiating behavior would be more evident when $n>3.0\times10^{11}$ /cm$^2$ where background impurities are no longer expected to have a significant impact on the experimentally measured mobility for our samples, and it could be useful to study this regime in the future to obtain further insight on whether $n_{RI}$ is constant or variable as a function of the 2DES density.

\subsection{3. Interface roughness scattering}	
	
	While a charged defect is the most straightforward scattering source for electrons in a GaAs 2DES, it is by no means the only one. Another, more subtle, yet potentially important mechanism to consider is interface roughness scattering. Layer fluctuations in the structure near the GaAs/AlGaAs barrier interface cause the electron energy level and charge distribution to experience a sharp change locally in the plane of the 2DES, generating a scattering potential for electrons. This principle applies to both single-interface and square-type QW structures, but the specifics of the formulations that are required to obtain scattering times are slightly different for the two situations. Since the majority of ultra-high-quality GaAs samples, including those used in this study, implement double-sided-doped QWs, here we follow the relatively simple formulation for QW structures with a finite barrier height \cite{Li}:
	
 \begin{equation}
\begin{split}
   &\frac{1}{\tau_{IR}}=\frac{4\pi m^* E_0^2 \Delta^2 \Lambda^2}{\hbar^3(L+\sqrt{\frac{2\hbar^2}{m^*(V_0-E_0)}})^2}f(\Lambda,k_F),
 \\
   &f(\Lambda,k_F)=\frac{1}{2\pi k_F^3}\int_{0}^{2k_F}(\frac{q}{q+q_{TF}})^2 \frac{e^{(-\frac{\Lambda^2 q^2}{4})}q^2\,dq}{\sqrt{1-(q/2k_F)^2}},
\label{eqn:5}
\end{split}
 \end{equation}
where $E_0$ is the ground-state energy of the QW, $V_0$ is the barrier height at the GaAs/AlGaAs interface, and $\Delta$, $\Lambda$ are the length scales of the fluctuations in the out-of-plane and in-plane directions of the 2DES, respectively. Although it is difficult to know the exact values for the roughness parameters, we can make rough estimates based on the fact that our samples are grown layer by layer using MBE. Given that one monolayer (ML) of GaAs is 2.83 {\AA} thick, it seems reasonable to postulate that $\Delta$ should be integer multiples of this thickness value. At our typical growth conditions of a monolayer per second with a substrate temperature of $T\simeq640$ \textdegree C, we speculate that $\Delta$ would be less than two monolayers. Throughout the rest of the paper, we assume that $\Delta\leq5.67$ {\AA} and only evaluate interface roughness scattering for the two cases of $\Delta=2.83$ {\AA} ($=1$ ML) and 5.67 {\AA} ($=2$ ML).
	
	The in-plane fluctuation length scale of $\Lambda$ is more difficult to assess solely from growth conditions. As described in Section VI, we evaluated the mobility of another set of samples with the QW width as the primary variable to address this issue specifically and estimate $\Lambda$ with some accuracy. We find that for our samples, there are two $\Lambda$ values each for both $\Delta=2.83$ ($\Lambda=30$ and 200 {\AA}) and $\Delta=5.67$ {\AA} ($\Lambda=34$ and 340 {\AA}) that yield scattering rates that match the data fairly well.  
	
	Figure 2(c) compares our experimental data with the $\mu$ vs. $n$ trends expected from Eq. (5) based on the four possible $\Delta$ and $\Lambda$ combinations (IR$_1$, IR$_2$, IR$_3$, and IR$_4$) outlined in the previous paragraph. The mobility we deduce for background impurities is also plotted, and the aggregate effect of $\tau_{BI}$ and $\tau_{IR}$ is shown with dashed curves of corresponding color. In Fig. 2(c) we do not include the effects of remote ionized impurity scattering to better visualize and understand the influence of $\tau_{IR}$ alone on our samples. 
	
	It is clear from Fig. 2(c) that when $\Lambda\geq200$ {\AA} (IR$_1$ and IR$_2$), $\tau_{IR}$ has a relatively insignificant consequence on the expected mobility of samples with background impurity concentrations similar to ours. The other two cases (IR$_3$ and IR$_4$) with $\Delta$ and $\Lambda$ being 2.83, 80 and 5.67, 34 {\AA}, respectively, appear to impact the mobility to a much larger extent. However, even in this situation it seems that additional contributions from other scattering terms are necessary to explain the experimental $\mu$ vs. $n$ data. While it may be tempting to consider completely neglecting $\tau_{IR}$ based on Fig. 2(c), it is important to remember that fluctuations on the monolayer level are difficult to avoid in the MBE growth of GaAs/AlGaAs samples. This implies that, although $\tau_{IR}$ could potentially have only a small effect on the total mobility of a GaAs 2DES, it should always be quantitatively assessed before being disregarded. In fact, we will see in Section VI that interface roughness scattering is essential in understanding the QW width dependence of mobility in our samples.
	
\section{VI. Mobility vs. Quantum well width}

	As discussed in the previous section, the primary difficulty in estimating $\tau_{IR}$ is the uncertainty of the lateral fluctuation length scale $\Lambda$. For GaAs 2DESs confined to square QWs it has been shown that interface roughness scattering dominates when the well width $w$ is small and gradually tapers off as $w$ increases \cite{Sakaki,Doby,Li}. One approach then to determine $\Lambda$ is to study the evolution of $\mu$ vs. $w$ in samples that have similar growth conditions to our ultra-high-quality GaAs 2DESs. Figures 3(a) and (c) show the measured mobility values for GaAs samples with $15\leq w\leq45$ nm and electron densities $n=1.2\times10^{11}$ and $1.7\times10^{11}$ /cm$^2$, respectively. The black, red, and blue symbols denote results from structures with different barrier alloy fractions of $x=0.12$, 0.24, and 0.36. For all cases, there is a clear trend of significant initial increase in $\mu$ when $w\leq30$ nm and then a slow saturation as $w$ increases further, even though impurity concentrations are nominally the same in these structures. Such behavior strongly suggests that $\tau_{IR}$ is an important contributor to the mobility, especially at narrow QW widths.
	
\begin{figure}[t]

\centering
    \includegraphics[width=.49\textwidth]{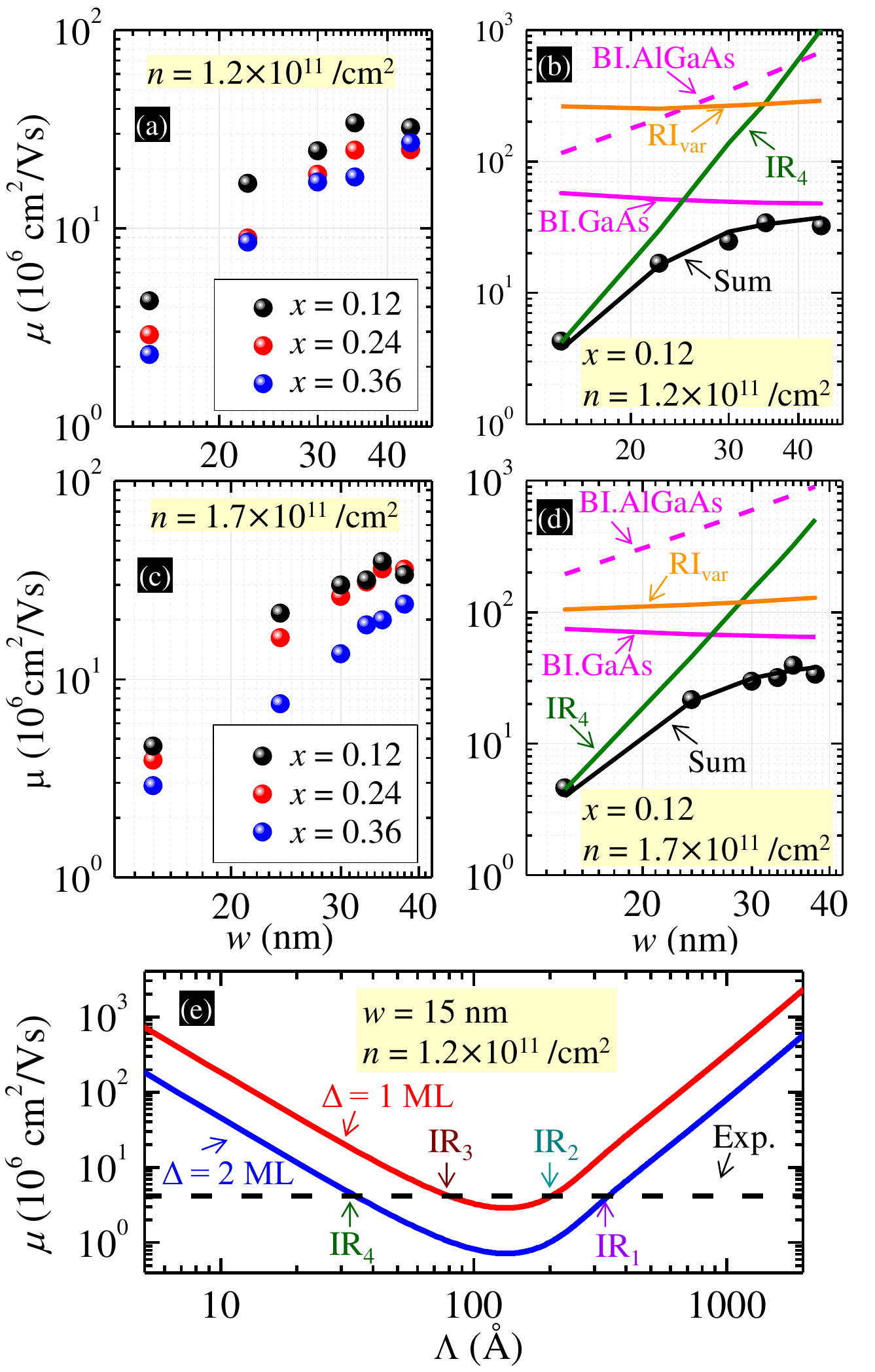} 
  \caption{\label{fig3} (a) $\mu$ vs. $w$ for GaAs 2DESs with $n=1.2\times10^{11}$ /cm$^2$. The black, red, and blue circles are for samples with $x=0.12$, 0.24, and 0.32 barriers, respectively. (b) Comparison of the $\mu$ vs. $w$ data for $x=0.12$ with values deduced from our models. The magenta solid and dashed curves represent scattering from $\tau_{BI.GaAs}$ and $\tau_{BI.AlGaAs}$, the orange solid curve is for $\tau_{RI}$ assuming $n_{RI}=n$, and the green solid curve is for $\tau_{IR}$ assuming the case IR$_4$. The solid black curve represents the total $\mu$ vs. $w$ trend deduced from Matthiessen's rule. (c) and (d) show similar plots for $n=1.7\times10^{11}$ /cm$^2$. (e) Expected $\mu$ vs. $\Lambda$ when only considering $\tau_{IR}$ for a sample with $w=15$ nm and $n=1.2\times10^{11}$ /cm$^2$. The dashed black line marks the experimental $\mu$, and the red and blue solid curves are for $\Delta=2.83$ {\AA} (1 ML) and 5.67 {\AA} (2 ML), respectively.}
\end{figure} 
	
	Figures 3(b) and (d) compare the $\mu$ vs. $w$ values deduced from considering all the scattering mechanisms discussed in Section V to the experimental data for the two cases of $n=1.2\times10^{11}$ and $1.7\times10^{11}$ /cm$^2$ when $x=0.12$. In our models, increasing $w$ can be qualitatively thought as replacing a finite amount of AlGaAs barrier with GaAs when considering $\tau_{BI}$. Similarly, a larger $w$ effectively pushes remote ionized dopants slightly further away from the 2DES when considering $\tau_{RI}$. The calculated results based on Eqs. 2-4 using $n_{BI.GaAs}=1.25\times10^{13}$ /cm$^3$ and $n_{BI.AlGaAs}=2.5\times10^{13}$ /cm$^3$ show that in the range of our study, impurity scattering is then only altered slightly as a function of $w$. This cannot explain the drastic drop in mobility observed in the data of narrow well samples. Given that when $w=15$ nm the measured $\mu$ is more than an order of magnitude smaller than the lowest value expected from impurity scattering for both densities, we assume that $\tau\simeq\tau_{IR}$ for this QW width and estimate $\Lambda$ using the two $\Delta$ values of 2.83 and 5.67 {\AA} as discussed earlier. 
	
	Based on this approach to narrow down $\Lambda$, we find four combinations of [$\Delta$, $\Lambda$] that match the data for $w=15$ nm and $n=1.2\times10^{11}$ /cm$^2$ when $x=0.12$ (Fig. 3(b)); [5.67, 340] (denoted IR$_1$), [2.83, 200] (IR$_2$), [2.83, 80] (IR$_3$), and [5.67, 34] (IR$_4$), all in units of {\AA}. (Our reasoning for why there are four pairs of [$\Delta$, $\Lambda$] values is based on Fig. 3(e) which we will discuss at the end of this section). While only the mobility expected from the combination IR$_4$ is plotted in Fig. 3(b), the other cases produce results that are indistinguishable over the entire covered range of $w$. Including all forms of impurity scattering along with the interface roughness scattering contributions to mobility shown in Fig. 3(b), we find excellent agreement between the collective results from our model and the experimental data. Using the roughness parameters for IR$_4$, we also find reasonably good agreement between our models and the $\mu$ vs. $w$ data for $1.7\times10^{11}$ /cm$^2$, as displayed in Fig. 3(d). Although the modeling results are only presented here for the $x=0.12$ case since this barrier condition is used for our ultra-high-quality samples, a similar procedure can be performed for the other barrier alloy fraction data sets in Figs. 3(a) and (c) to obtain roughness parameters for different types of structures.  Assuming that the change in $x$ does not significantly alter impurity scattering contributions from $\tau_{BI.AlGaAs}$, this would encompass modifying the barrier height $V_0$ and corresponding change in ground-state energy $E_0$ in Eq. 5 to deduce $\tau_{IR}$ and compare the results with the mobility for the $w=15$ nm samples with $x=0.24$ and 0.36.

	It is possible to evaluate $\Lambda$ by matching the interface roughness scattering to the $\mu$ vs. $w$ data for $1.7\times10^{11}$ /cm$^2$ rather than $1.2\times10^{11}$ /cm$^2$. The four [$\Delta$, $\Lambda$] combinations that correspond to IR$_1$ to IR$_4$ for this higher density are [5.67, 282], [2.83, 163], [2.83, 72], and [5.67, 30], all in units of {\AA}. In the higher density regime where $\tau_{IR}$ starts to become more important in our samples, we find that these parameters imply less interface roughness scattering compared to earlier. This indicates that while the resultant $\Lambda$ values may be slightly different from what was discussed in the previous paragraph when we optimize them for a different set of samples, they do not alter the conclusions we made earlier based on the $\mu$ vs. $n$ data in Fig. 2(c). 
	
	We now remark on why there are multiple [$\Delta$, $\Lambda$] combinations that yield very similar interface roughness scattering rates. Figure 3(e) shows the expected mobility of a sample with $w=15$ nm and $n=1.2\times10^{11}$ /cm$^2$ only considering $\tau_{IR}$ as a function of $\Lambda$ for the two cases of $\Delta=2.83$ (red) and 5.67 {\AA} (blue solid curve). The dashed line marks the mobility we measure experimentally. The trend of $\mu$ vs. $\Lambda$ repeats for both $\Delta$ values, showing an initial decline and then increase with a minimum at $\Lambda\simeq130$ {\AA}. Such a behavior has also been observed in other studies of interface roughness scattering \cite{Li}. This is somewhat puzzling at first sight if we recall that $\Lambda$ is the length scale of layer fluctuations in the plane of the 2DES, since naively one would expect a monotonous improvement in mobility as the interface becomes smoother and $\Lambda$ increases. However, it is important to bear in mind that in a simple zero-(or very-low-)temperature model, electron scattering only occurs near the Fermi level. For the case of Fig. 3(e), $k_F=8.7\times10^7$ /m, which corresponds to a characteristic length of $\lambda\sim700$ {\AA}. When $\Lambda<<\lambda$, traversing electrons are impervious to roughness because it averages out over the length of $\lambda$. The layer fluctuations scatter electrons most violently as $\Lambda$ increases and becomes comparable to $\lambda$, and then the effect dies out as $\Lambda$ increases further and the roughness becomes `transparent' to electrons again. 
	
\section{VII. Limits to mobility in current ultra-high-quality GaAs 2DES\MakeLowercase{s}}

	Based on what was established in the previous sections, we are now equipped with plausible parameters for our structures to consider all the scattering mechanisms that are relevant for our ultra-high-quality GaAs 2DESs. The next step is to put a comprehensive picture together and compare the results with the experimental data so that we can understand what limits mobility in our state-of-the-art GaAs samples. As briefly discussed in Section V.1, in the low-density regime where $n\lesssim1.6\times10^{11}$/cm$^2$, we expect $\tau_{IR}$, $\tau_{RI}>>\tau_{BI}$ so that background impurity scattering determines the mobility. To obtain reasonably good agreement between the computed results and experimental data for $\mu$ vs. $n$ in this density range, it is necessary to keep $n_{BI.GaAs}$ within $\sim10$\% variance of the $1.25\times10^{13}$ /cm$^3$ estimated earlier. While there is a larger margin of error for $n_{BI.AlGaAs}$, we find that it should still be less than $5\times10^{13}$ /cm$^3$ for $\tau_{BI}$ to agree well with the experimental data. We therefore believe that the credibility of these numbers is fairly high and keep them fixed when deducing mobility from the holistic scattering rate in the following paragraphs.
	
	
\begin{figure*}[t]
 
 \centering
    \includegraphics[width=.90\textwidth]{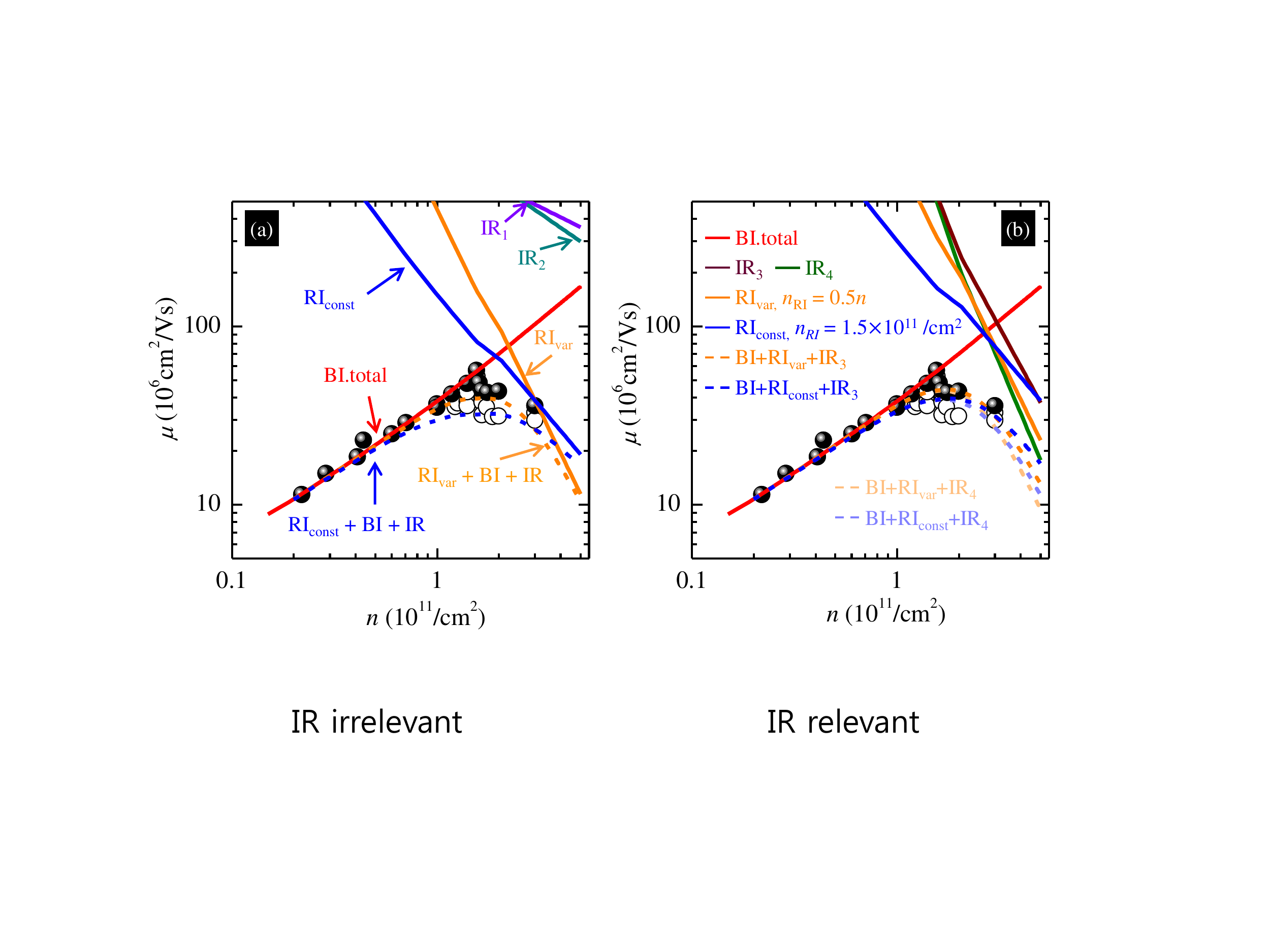} 
  \caption{\label{fig4} Interpreting our $\mu$ vs. $n$ data after considering all scattering mechanisms. (a) shows the scenario where $\tau_{IR}$ is very large, such as in the case when $\Delta=5.67$ {\AA}, $\Lambda=340$ {\AA} (IR$_1$, dark red solid line), or $\Delta=2.83$ {\AA}, $\Lambda=200$ {\AA} (IR$_2$, cyan solid line), and is almost irrelevant in determining the total scattering rate. As in the case of Fig. 2(b), the solid red line shows the $\mu$ vs. $n$ trend deduced from $\tau_{BI}$, while the blue and orange solid lines come from $\tau_{RI}$ assuming $n_{RI}=3.0\times10^{11}$ /cm$^2$ and $n_{RI}=n$, respectively. The dashed blue and orange lines show the total $\mu$ calculated from Matthiessen's rule considering $\tau_{BI}$, $\tau_{RI}$, and $\tau_{IR}$, with the colors representing the $\tau_{RI}$ condition used in the model. The two different $\tau_{IR}$ trends here have $\mu$ values that are too large to make a difference in the total $\mu$, which is almost fully determined by $\tau_{BI}$ and $\tau_{RI}$ as in Fig. 2(b). (b) shows a different scenario where $\tau_{IR}$ is chosen to be comparable to the other scattering rates, such as in the case when $\Delta=2.83$ {\AA}, $\Lambda=80$ {\AA} (IR$_3$, purple solid line), or $\Delta=5.67$ {\AA}, $\Lambda=34$ {\AA} (IR$_4$, green solid line). Here the solid blue and orange lines also represent $\mu$ vs. $n$ trends deduced from $\tau_{RI}$, but with lower impurity concentrations of $n_{RI}=1.5\times10^{11}$ /cm$^2$ and $n_{RI}=0.5n$ compared to Fig. 2(b). The dashed lines show the total $\mu$ vs. $n$ trends deduced from Matthiessen's rule, with the combination of scattering terms used denoted in the legend of the figure for each color.}
\end{figure*} 

	The implications of our model are not as clearcut for the samples with higher density ($n\gtrsim1.6\times10^{11}$ /cm$^2$). The multiple options available for $\tau_{RI}$ and $\tau_{IR}$ present several different explanations for the data, as shown in Figs. 2(b), 2(c), and 4. We reiterate first that, if we assume that interface roughness scattering has negligible effect and is essentially irrelevant, then the best fits to our data are those shown by dashed lines in Fig. 2(b). These fits are reasonable, but they do miss the data points with the highest mobility values. On the other hand, if we assume that remote ionized scattering is irrelevant and that the dominant scattering at high densities is via interface roughness, we obtain the dashed lines in Fig. 2(c). None of these curves are good fits through the highest-density data points, but they do match the highest-mobility data points up to $n\simeq1.6\times10^{11}$ /cm$^2$ very well.

	Now, is there a combination of interface roughness and remote impurity scattering mechanisms that could explain the data better? The short answer is no. To demonstrate this point, in Fig. 4(a) we show the results if we assume that scattering by remote impurities is present and that either IR$_1$ or IR$_2$ are the relevant interface roughness scattering parameters. (Recall that, based on our analysis described in Section VI, all four combinations of interface roughness scattering, IR$_1$, IR$_2$, IR$_3$, and IR$_4$, explain the dependence of mobility on QW width equally well.) Consistent with what can be expected from Fig. 2(c) for IR$_1$ and IR$_2$, in Fig. 4(a) $\tau_{IR}$ only starts to become relevant when $\mu>10^8$ cm$^2$/Vs and the results we obtain from Matthiessen's rule are almost identical to those shown in Fig. 2(b). In this scenario, the mobility of our high-density, ultra-high-quality GaAs 2DESs is most strongly determined by remote ionized impurity scattering but note that the fits, which include interface roughness scattering, are not better than Fig. 2(b) fits. 

	Alternatively, it could be that interface roughness scattering does play a substantial role in deciding the total mobility, and that IR$_3$ or IR$_4$ are the relevant parameters. The best fits to the experimental data in this case are shown in Fig. 4(b). Note that, for reasonable fits, we would have to assume smaller values for remote ionized impurity concentrations, namely, $n_{RI}=1.5\times10^{11}$ /cm$^2$ or $n_{RI}=0.5n$; these are each a factor of two smaller than those used in Fig. 2(b). The overall fits for the total mobility are comparable to those seen in Figs. 2(b), 2(c), and 4(a). Note that these assumptions for $n_{RI}$ imply that $n_{RI}$ is smaller than the density of electrons in the 2DES. This may sound implausible. However, it has been suggested that for the doping-well structure used in our samples as well as other ultra-high-quality GaAs 2DESs \cite{Manfra35,HighMobility,Dopingwell}, an additional screening term associated with excess electrons in the doped region should be included for the analysis of charged-impurity scattering \cite{Boris1,Boris2}. This is especially true for $\tau_{RI}$ because the excess electrons enable correlation between the remote ionized impurities, resulting in a structure-factor-based reduction in scattering. At the most basic level, such extra screening can be viewed as having a similar effect as decreasing $n_{RI}$ to a smaller effective value in Eq. (4). Using this crude logic, allowing $n_{RI}<n$ in our models to reduce the influence of $\tau_{RI}$ on the total mobility could be justified.

	In both Figs. 4(a) and (b), from our models and analyses, it appears that remote ionized impurity scattering has an impact on the mobility in the high-density regime. This seems consistent with empirical anecdotes that the peak mobility of ultra-high-quality GaAs 2DESs is often sensitive to the cooldown or illumination procedures when measured in a given sample. There is not much room for $\tau_{IR}$ to change after growth, but $\tau_{RI}$ can depend on the specific measurement details because the excess screening conditions mentioned in the previous paragraph could vary for different cooldown and illumination procedures \cite{Boris1,Boris2}. For example, if the sample is cooled down too fast, it may not give the excess electrons in the doped region enough time to arrange themselves in an optimal orientation and therefore reduce the amount of screening provided.

	Given the above understandings, we now discuss possible avenues to further improve the mobility in GaAs 2DESs. It is clear that mobility is limited by background impurities when $n\lesssim1.6\times10^{11}$ /cm$^2$ and by remote ionized impurities and possibly interface rougness at higher densities. We believe there are two approaches going forward. The first approach would be to continue to reduce the concentration of background impurities by improving the vacuum in the MBE chamber and purify the source Ga, Al, and As source materials. While this is conceptually straightforward, it is extremely challenging, given the excruciating efforts that are necessary \cite{HighMobility}. Perhaps a more feasible strategy is to reduce the influence of remote ionized impurities by modifying the sample design. For example, instead of decreasing the spacer thickness to increase $n$, one could start off with an optimized, low-density sample and apply a positive voltage bias to a gate on the top or bottom side of the sample. Using this technique we project that, in principle, one would be able to achieve $\mu=100\times10^6$ cm$^2$/Vs at $n\simeq3\times10^{11}$ /cm$^2$, assuming that the sample has a sufficiently small QW width (to avoid the occupation of the second electric subband), a sufficiently thick spacer, and that it can withstand large biases. However, it is important to consider that single-sided gating would alter the charge distribution, possibly causing additional interface roughness scattering and decrease $\tau_{IR}$ as the electron wavefunction is pressed against the GaAs/AlGaAs interface. Symmetric gating of the sample is therefore desirable, but this requires the deposition of a front gate which could degrade the mobility of the sample during processing. Even in the ideal case, where gating from both sides is achieved, it is still possible that interface roughness scattering would eventually take over and limit the mobility enhancement. If this is the case, it would be necessary to develop growth processes, such as judicious growth interruptions, that optimize $\Delta$ and $\Lambda$ to suppress the influence of $\tau_{IR}$. 


\begin{acknowledgments}
We acknowledge support by the National Science Foundation (NSF) Grant Nos. DMR 2104771, ECCS 1906253, and MRSEC DMR 2011750, the Eric and Wendy Schmidt Transformative Technology Fund, and the Gordon and Betty Moore Foundations EPiQS Initiative (Grant No. GBMF9615 to L.N.P.) for sample fabrication and characterization. For measurements, we acknowledge support by the U.S. Department of Energy Basic Energy Sciences (Grant No. DEFG02-00-ER45841). M.S. acknowledges partial support through a QuantEmX travel grant from the Institute for Complex Adaptive Matter and the Gordon and Betty Moore Foundation through Award No. GBMF5305. We also thank Christian Reichl and Werner Wegscheider for illuminating discussions.
 \end{acknowledgments}

 The data that support the findings of this study are available from the corresponding author upon reasonable request.


\begin{thebibliography}{99}

\bibitem{Tsui} D. C. Tsui, H. L. Stormer, and A. C. Gossard, Two-dimensional magnetotransport in the extreme quantum limit. Phys. Rev. Lett. {\bf 48}, 1559 (1982).

\bibitem{Willett} R. Willett, J. P. Eisenstein, H. L. St\"{o}rmer, D. C. Tsui, A. C. Gossard, and J. H. English, Observation of an even-denominator quantum number in the fractional quantum Hall effect. Phys. Rev. Lett. {\bf 59}, 1776 (1987).

\bibitem{stripe1} M. P. Lilly, K. B. Cooper, J. P. Eisenstein, L. N. Pfeiffer, and K. W. West, Evidence for an anisotropic State of two-dimensional electrons in high Landau levels. Phys. Rev. Lett. {\bf 82}, 394 (1999).

\bibitem{stripe2} R. R. Du, D. C. Tsui, H. L. Stormer, L. N. Pfeiffer, K. W. Baldwin, and K. W. West, Strongly anisotropic transport in higher two-dimensional Landau levels. Solid State Commun. {\bf 109}, 389 (1999).

\bibitem{HeWigner} C. C. Grimes and G. Adams, Evidence for a liquid-to-crystal phase transition in a classical, two-dimensional sheet of electrons. Phys. Rev. Lett. {\bf 42}, 795 (1979).

\bibitem{Wigner1} E. Y. Andrei, G. Deville, D. C. Glattli, F. I. B. Williams, E. Paris, and B. Etienne, Observation of a magnetically induced Wigner solid. Phys. Rev. Lett. {\bf 60}, 2765 (1988).

\bibitem{Wigner2} H. W. Jiang, R. L. Willett, H. L. St\"{o}rmer, D. C. Tsui, L. N. Pfeiffer, and K. W. West, Quantum liquid versus electron solid around $\nu=1/5$ Landau-level filling. Phys. Rev. Lett. {\bf 65}, 633 (1990).

\bibitem{Wigner3} V. J. Goldman, M. Santos, M. Shayegan, and J. E. Cunningham, Evidence for two-dimensional quantum Wigner crystal. Phys. Rev. Lett. {\bf 65}, 2189 (1990).

\bibitem{B1} M. Kellogg, J. P. Eisenstein, L. N. Pfeiffer, and K. W. West, Vanishing Hall resistance at high magnetic field in a double-layer two-dimensional electron system. Phys. Rev. Lett. {\bf 93}, 036801 (2004).

\bibitem{B2} E. Tutuc, M. Shayegan, and D. A. Huse, Counterflow measurements in strongly correlated GaAs hole bilayers: Evidence for electron-hole pairing. Phys. Rev. Lett. {\bf 93}, 036802 (2004).

\bibitem{B3} J. P. Eisenstein and A. H. MacDonald, Bose-Einstein condensation of excitons in bilayer electron systems. Nature {\bf 432}, 691 (2004).

\bibitem{ShayeganReview} M. Shayegan, Flatland electrons in high magnetic fields. in {\it High Magnetic Fields: Science and Technology}, edited by F. Herlach and N. Miura (World Scientific, Singapore, 2006), Vol. 3, pp. 31-60.

\bibitem{JainBook} J. K. Jain, {\it Composite Fermions} (Cambridge University Press, Cambridge, England, 2007).

\bibitem{HalperinBook} B. I. Halperin, J. K. Jain, {\it Fractional Quantum Hall effects: New developments} (World Scientific, Singapore, 2020).

\bibitem{modulation} R. Dingle, H. L. St\"{o}rmer, A. C. Gossard, and W. Wiegmann, Electron mobilities in modulation-doped semiconductor heterojunction superlattices. Appl. Phys. Lett. {\bf 33}, 665 (1978).

\bibitem{EnglishM} J. H. English, A. C. Gossard, H. L. St\"{o}rmer, and K. W. Baldwin, GaAs structures with electron mobility of $5\times10^6$ cm$^2$/Vs. Appl. Phys. Lett. {\bf 50}, 1826 (1987).

\bibitem{ParisM} B. Etienne and E. Paris, Two-dimensional electron gas of very high mobility in planar doped heterostructures. J. Phys. France {\bf 48}, 2049 (1987).

\bibitem{ShayeganM} M. Shayegan, V. J. Goldman, C. Jiang, T. Sajoto, and M. Santos, Growth of low-density two-dimensional electron system with very high mobility by molecular beam epitaxy. Appl. Phys. Lett. {\bf 52}, 1086 (1988).

\bibitem{ShayeganM2} M. Shayegan, V. J. Goldman, M. Santos, T. Sajoto, L. Engel, and D. C. Tsui, Two-dimensional electron system with extremely low disorder. Appl. Phys. Lett. {\bf 53}, 2080 (1988).

\bibitem{PfeifferM} Loren Pfeiffer, K. W. West, H. L. Stormer, and K. W. Baldwin, Electron mobilities exceeding $10^7$ cm$^2$/Vs in modulation-doped GaAs. Appl. Phys. Lett. {\bf 55}, 1888 (1989).

\bibitem{Stormer} H. L. St\"{o}rmer, R. Dingle, A. C. Gossard, W. Wiegmann, and M. D. Sturge, Two-dimensional electron gas at a semiconductor-semiconductor interface. Solid State Commun. {\bf 88}, 933 (1993).

\bibitem{LorenPhysica} L. Pfeiffer and K. W. West, The role of MBE in recent quantum Hall effect physics discoveries. Physica E {\bf 20}, 57 (2003).

\bibitem{Umansky35} V. Umansky, M. Heiblum, Y. Levinson, J. Smet, J. N{\"u}bler, and M. Dolev, MBE growth of ultra-low disorder 2DEG with mobility exceeding $35\times10^6$ cm$^2$/Vs. J. Cryst. Growth {\bf 311}, 1658 (2009).

\bibitem{Schlom} D. G. Schlom and L. N. Pfeiffer, Upward mobility rocks! Nat. Mater. {\bf 9}, 881 (2010).

\bibitem{Manfra35} M. J. Manfra, Molecular beam epitaxy of ultra-high-quality AlGaAs/GaAs heterostructures: Enabling physics in low-dimensional electronic systems. Annu. Rev. Condens. Matter Phys. {\bf 5}, 347 (2014).

\bibitem{Alproblem} Y. J. Chung, K. W. Baldwin, K. W. West, M. Shayegan, and L. N. Pfeiffer, Surface segregation and the Al problem in GaAs quantum wells. Phys. Rev. Mater. {\bf 2}, 034006 (2018).

\bibitem{Dopingwell} Y. J. Chung, K. A. Villegas Rosales, K. W. Baldwin, K. W. West, M. Shayegan, and L. N. Pfeiffer, Working principles of doping-well structures for high-mobility two-dimensional electron systems. Phys. Rev. Mater. {\bf 4}, 044003 (2020).

\bibitem{WegS} E. K\"{u}lah, C. Reichl, J. Scharnetzky, L. Alt, W. Dietsche, and W. Wegscheider, The improved inverted AlGaAs/GaAs
interface: its relevance for high-mobility quantum wells and hybrid systems. Semicond. Sci. Technol. {\bf 36}, 085013 (2021).

\bibitem{HighMobility} Y. J. Chung, K. A. Villegas-Rosales, K. W. Baldwin, P. T. Madathil, K. W. West, M. Shayegan, and L. N. Pfeiffer, Ulta-high-quality two-dimensional electron systems. Nat. Mater. {\bf 20}, 632 (2021).

\bibitem{HighHole} Y. J. Chung, C. Wang, S. K. Singh, A. Gupta, K. W. Baldwin, K. W. West, M. Shayegan, L. N. Pfeiffer, and R. Winkler, Record-quality GaAs two-dimensional hole systems. Phys. Rev. Mater. {\bf 6}, 034005 (2022).

\bibitem{Jiang} C. Jiang, D. C. Tsui, and G. Weimann, Threshold transport of high-mobility two-dimensional electron gas in GaAs/AlGaAs heterostructures. Appl. Phys. Lett. {\bf 53}, 1533 (1988).

\bibitem{Gold} A. Gold, Mobility of the two-dimensional electron gas in AlGaAs/GaAs heterostructures at low electron densities. Appl. Phys. Lett. {\bf 54}, 2100 (1989).

\bibitem{Gold2} A. Gold, Temperature dependence of mobility in Al$_x$Ga$_{1-x}$As/GaAs heterostructures for impurity scattering. Phys. Rev. B {\bf 41}, 8537 (1990).

\bibitem{DS1} E. H. Hwang and S. Das Sarma, Limit to two-dimensional mobility in modulation-doped GaAs quantum structures: How to achieve a mobility of 100 million. Phys. Rev. B {\bf 77}, 235437 (2008).

\bibitem{DS2} S. Das Sarma, E. H. Hwang, S. Kodiyalam, L. N. Pfeiffer, and K. W. West, Transport in two-dimensional modulation-doped semiconductor structures. Phys. Rev. B {\bf 91}, 205304 (2015).

\bibitem{Boris1} M. Sammon, M. A. Zudov, and B. I. Shklovskii, Mobility and quantum mobility of modern GaAs/AlGaAs heterostructures. Phys. Rev. Materials {\bf 2}, 064604 (2018).

\bibitem{Boris2} M. Sammon, T. Chen, and B. I. Shklovskii, Excess electron screening of remote donors and mobility in modern GaAs/AlGaAs heterostructures. Phys. Rev. Materials {\bf 2}, 104001 (2018).

\bibitem{DS3} S. Ahn and S. Das Sarma, Density-dependent two-dimensional optimal mobility in ultra-high-quality semiconductor quantum wells. Phys. Rev. Mater. {\bf 6}, 014603 (2022).

\bibitem{Boris3} Yi Huang, B. I. Shklovskii, and M. A. Zudov, Scattering mechanisms in state-of-the-art GaAs/AlGaAs quantum wells. cond-mat.mes-hall arXiv:2205.00365

\bibitem{Delta} C. E. C. Wood, G. Metze, J. Berry, and L. F. Eastman, Complex free carrier profile synthesis by ``atomic-plane'' doping of MBE GaAs. J. Appl. Phys. {\bf 51}, 383 (1980).

\bibitem{Ploog} K. Ploog, Delta-($\delta$-) doping in MBE-grown GaAs: Concept and device application. J. Cryst. Growth {\bf 81}, 304 (1987).

\bibitem{Si1} M. Santos, T. Sajoto, A. Zrenner, and M. Shayegan, Effect of substrate temperature on migration of Si in planar doped GaAs. Appl. Phys. Lett. {\bf 53}, 2504 (1988); Erratum in Appl. Phys. Lett. {\bf 55}, 603 (1989).

\bibitem{Si2} A-M. Lanzillotto, M. Santos, and M. Shayegan, Secondary-ion mass spectrometry study of the migration of Si in planar-doped GaAs and Al$_{0.25}$Ga$_{0.75}$As. Appl. Phys. Lett. {\bf 55}, 1445 (1989).

\bibitem{Si3} A-M. Lanzillotto, M. Santos, and M. Shayegan, Silicon migration during the molecular beam epitaxy of delta-doped GaAs and Al$_{0.25}$Ga$_{0.75}$As. J. Vac. Sci. Technol. A {\bf 8}, 2009 (1990).

\bibitem{Schubert} E. F. Schubert, Delta doping of III-V compound semiconductors: Fundamentals and device applications. J. Vac. Sci. Technol. A {\bf 8}, 2980 (1990).

\bibitem{Davies} J. H. Davies, \textit{The physics of low dimensional semiconductors}, (Cambridge University Press, Cambridge, 1997).

\bibitem{Designrules} Y. J. Chung, K. W. Baldwin, K. W. West, D. Kamburov, M. Shayegan, and L. N. Pfeiffer, Design rules for modulation-doped AlAs quantum wells, Phys. Rev. Mater. {\bf 1}, 021002(R) (2017).

\bibitem{Ourmazd} A. Ourmazd, D. W. Taylor, J. Cunningham, and C. W. Tu, Chemical mapping of semiconductor interfaces at near-atomic resolution. Phys. Rev. Lett. {\bf 62}, 933 (1989).

\bibitem{Salemink} H. W. M. Salemink and O. Albrektsen, Atomic-scale composition fluctuations in III-V semiconductor alloys. Phys. Rev. B {\bf 47}, 16044(R) (1993).

\bibitem{Zrenner} A. Zrenner, L. V. Butov, M. Hagn, G. Abstreiter, G. B{\"o}hm, and G. Weimann, Quantum dots formed by interface fluctuations in AlAs/GaAs coupled quantum well structures. Phys. Rev. Lett. {\bf 72}, 3382 (1994).

\bibitem{Ando} T. Ando, A. B. Fowler, F. Stern, Electronic properties of two-dimensional systems. Rev. Mod. Phys. {\bf 54}, 437 (1982).

\bibitem{GoldIR1} A. Gold, Electronic transport properties of a two-dimensional electron gas in a silicon quantum-well structure at low temperature. Phys. Rev. B {\bf 35}, 723 (1987).

\bibitem{GoldIR2} A. Gold, Scattering time and single-particle relaxation time in a disordered two-dimensional electron gas. Phys. Rev. B {\bf 38}, 10798 (1988).

\bibitem{Li} J. M. Li, J. J. Wu, X. X. Han, Y. W. Lu, X. L. Liu, Q. S. Zhu, and Z. G. Wang, A model for scattering due to interface roughness in finite quantum wells. Semicond. Sci. Technol. {\bf 20}, 1207 (2005).

\bibitem{Sakaki} H. Sakaki, T. Noda, K. Hirakawa, M. Tanaka, and T. Matsusue, Interface roughness scattering in GaAs/AlAs quantum wells. Appl. Phys. Lett. {\bf 51}, 1934 (1987).

\bibitem{Doby} D. Kamburov, K. W. Baldwin, K. W. West, M. Shayegan, and L. N. Pfeiffer, Interplay between quantum well width and interface roughness for electron transport mobility in GaAs quantum wells. Appl. Phys. Lett. {\bf 109}, 232105 (2016).


\end{thebibliography}
\end{document}